\documentclass[sigconf]{acmart}

\AtBeginDocument{%
  \providecommand\BibTeX{{%
    \normalfont B\kern-0.5em{\scshape i\kern-0.25em b}\kern-0.8em\TeX}}}
    
\usepackage{booktabs} 
\usepackage{balance} 

\usepackage{amssymb,amsmath}
\usepackage{xcolor}
\usepackage{listings}
\usepackage{float}
\usepackage{tikz}
\usepackage{enumitem}
\usepackage{showexpl}
\usepackage{graphics}
\usepackage{rotating}
\usepackage{xspace}
\usepackage{colortbl}
\usepackage{tabularx}
\usepackage{multirow}
\usepackage{url}
\usepackage{booktabs}
\usepackage{color}
\usepackage{subfig}
\usepackage{algorithm}
\usepackage{algpseudocode}
\usepackage{hhline}

\usepackage{multicol}
\usepackage{textgreek}
\usepackage{bbold}

\usetikzlibrary{automata,shapes.multipart}
\usetikzlibrary[arrows,decorations.pathmorphing,backgrounds,positioning,fit,petri]
\usetikzlibrary{fit}

\newcommand{\naturals}{\mathbb{N}}
\newcommand{\update}{{\xi}}
\newcommand{\ts}{{\mathit{TS}}}

\newcommand{\D}{\mathbb{D}}
\renewcommand{\path}{\rho}

\newcommand{\one}{\mathbb{1}}
\newcommand{\zero}{\mathbb{0}}

\newcommand{\wb}{\mathit{WB}}
\newcommand{\sbbas}{\mathit{SB}}
\newcommand{\bas}{\mathit{bas}}

\newcommand{\true}{\mathit{true}}
\newcommand{\false}{\mathit{false}}

\newcommand{\hd}{\mathit{hd}}
\newcommand{\reach}{\mathit{reach}}

\newcolumntype{C}[1]{>{\centering\arraybackslash}p{#1}}
\newcolumntype{L}[1]{>{\raggedright\arraybackslash}p{#1}}
\newcolumntype{R}[1]{>{\raggedleft\arraybackslash}p{#1}}

\tikzset{
  bend angle=30,
  ->,
  shorten >=1pt,
  node distance=1.3cm and 1.3cm,
  on grid,
  auto,
  initial where=above,
  initial text=,
  initial distance=0.6cm,
  inner sep=0.5mm,
  smallstate/.style={circle,draw},
  rootstate/.style={rectangle,draw},
  openstate/.style={},
  succstate/.style={font={$\surd$}},
  bendanglelarge/.style={bend angle=45},
  bendanglesmall/.style={bend angle=10},
  dim/.style={lightgray},
  diag/.style={red},
  bisim/.style={red,dashed,-,bend left},
  loop above/.style={in=110,out=70,loop,distance=0.5cm},
  loop left/.style={in=200,out=160,loop,swap,distance=0.5cm,looseness=1},
  loop right/.style={in=20,out=-20,loop,distance=0.5cm,looseness=1},
  loop below/.style={in=290,out=250,loop,distance=0.5cm},
  attr/.style={draw,fill=black!15,rounded corners}
}

\definecolor{darkred}{rgb}{0.55, 0.0, 0.0}
\begin{document}
\title[Target Control of Boolean Networks]{A Dynamics-based Approach for \\ the Target Control of Boolean Networks}

\author{Cui Su}
\affiliation{
\institution{Interdisciplinary Centre for Security, Reliability and Trust, University of Luxembourg}
}
\author{Jun Pang}
\affiliation{
\institution{Interdisciplinary Centre for Security, Reliability and Trust \&
Faculty of Science, Technology and Medicine, University of Luxembourg}
}

\begin{abstract}
We study the target control problem of asynchronous Boolean networks, to identify a set of nodes, the perturbation of which can drive the dynamics of the network from any initial state to the desired steady state (or attractor). We are particularly interested in temporary perturbations, which are applied for sufficient time and then released to retrieve the original dynamics. Temporary perturbations have the apparent advantage of averting unforeseen consequences, which might be induced by permanent perturbations. Despite the infamous state-space explosion problem, in this work, we develop an efficient method to compute the temporary target control for a given target attractor of a Boolean network. We apply our method to a number of real-life biological networks and compare its performance with the stable motif-based control method to demonstrate its efficacy and efficiency.
\end{abstract}

\keywords{Boolean networks, attractors, network control}

\maketitle

\section{Introduction}\label{sec:intro}

Cell reprogramming has garnering attention for its therapeutic potential for treating the most devastating diseases characterised by diseased cells or a deficiency of certain cells. 
It is capable of reprogramming any kind of abundant cells in the body into the desired deficient cells to restore functions of the diseased organ~\citep{SD16,GD19,GMS19}. 
It has shown promising benefits for clinical applications, such as cell and tissue engineering, regenerative medicine and drug discovery.

In their seminal work, Yamanaka {\it et al.} showed that human somatic cells can be converted to induced pluripotent stem cells (iPSCs) by a cocktail of defined factors~\citep{Yama07,TTON07}. 
The generated iPSCs have the ability to further propagate and differentiate into many cell types. 
However, the application of iPSC reprogramming is often restricted, due to that: 
(1) the generated iPSCs have a risk of cancerous tumour formation~\citep{GMS19,GD19}; 
(2) the iPSC reprogramming and differentiation process usually requires long time to produce sufficient cells for application, which leads to a significant experimental cost~\citep{GD19}; and 
(3) the iPSCs often encounter cell cycle arrest after differentiation, which makes it impossible to expand the number of cells for therapeutic transplantation~\citep{GMS19}.
The limitations of iPSC reprogramming reinforce the need of direct reprogramming, 
also called transdifferentiation. 
Direct reprogramming harnesses abundant somatic cells to regenerate defective cells by reprogramming the somatic cells directly into the desired cell type bypassing the pluripotent state.
As a consequence, direct reprogramming can not only reduce the risk of tumourigenesis and teratoma formation, but also shorten the period of time for producing enough desired cells for therapeutic application.

A major challenge of cell reprogramming lies in the identification of effective target proteins or genes, the manipulation of which can trigger desired changes. 
Lengthy time commitment and high cost hinder the efficiency of experimental approaches, which perform brute-force tests of tunable parameters and record corresponding results~\citep{L16}.
This strongly motivates us to turn to mathematical modelling of biological systems, which allows us to identify key genes or pathways that can trigger desired changes using computational methods. 
Boolean network, first introduced by Kauffman~\citep{KS69}, is a well-established modelling framework for gene regulatory networks and their associated signalling pathways, and it has apparent advantages compared to other modelling frameworks~\citep{Aku18}. 
Boolean network provides a qualitative description of biological systems and thus evades the parametrisation problem, which often occurs in quantitative models, such as models of ordinary differential equations (ODEs). 
In Boolean networks, molecular species, such as genes and transcription factors, are described as Boolean variables. 
Each variable is assigned with a Boolean function, which determines the evolution of the node. 
Boolean functions characterise activation or inhibition regulations between molecular species. 
The dynamics of a Boolean network is assumed to evolve in discrete time steps, moving from one state to the next, under one of the updating schemes, such as {\it synchronous} or {\it asynchronous}. 
Under the synchronous scheme, all the nodes update their values simultaneously at each time step; while under the asynchronous scheme, only one node is randomly selected to update its value at each time step. 
We focus on the asynchronous updating scheme since it can capture the phenomenon that biological processes occur at different time scales. 
The steady-state behaviour of the dynamics is described as {\it attractors}, to one of which the system eventually settles down. 
Attractors are hypothesised to characterise cellular phenotypes~\citep{HS01}. 
Each attractor has a {\it weak basin} and a {\it strong basin}. 
The weak basin contains all the states that can reach this attractor, 
while the strong basin includes the states that can only reach this attractor and cannot reach any other attractors of the network. 
In the context of Boolean networks, cell reprogramming is interpreted as a control problem: modifying the parameters of a network to lead its dynamics towards a desired attractor.

Control theories have been employed to modulate the dynamics of complex networks in recent years. 
Due to the intrinsic non-linearity of biological systems, control methods designed for linear systems, 
such as structure-based control methods~\citep{LSB11,GLDB14,CGCK16}, are not applicable -- 
they can both overshoot and undershoot the number of control nodes for non-linear networks~\citep{GR16}. 
For nonlinear systems of ODEs, Fiedler {\it et al.} proved that the control of a feedback vertex set is sufficient to control the entire network~\citep{ABGD13,BAGD13,ZYA17}; 
and Cornelius {\it et al.} proposed a simulation-based method to predict instantaneous perturbations 
that can reprogram a cell from an undesired phenotype to a desired one.
However, further study is required to figure out if these two methods can be lifted to control Boolean networks. 
Several methods based on semi-tensor product (STP) have been proposed to solve different control problems for Boolean control networks (BCNs) under the synchronous updating scheme~\citep{LCL17,ZLLC18,LZHY16,ZLKS19,WSZS19,CLW16,YYCJ19,ZKF13}. 
For synchronous Boolean networks, Kim {\it et al.} developed a method to compute a small fraction of nodes, 
called `control kernels', that can be modulated to govern the dynamics of the network~\citep{KSK13};  and
Moradi {\it el al.} developed an algorithm guided by forward dynamic programming to solve the control problem. 
However, all these methods are not directly applicable to asynchronous Boolean networks.
To tackle this problem, 
we have developed several decomposition-based methods, which exploit both the structural and dynamical information, to cope with source-target control with instantaneous, temporary and permanent perturbations~\citep{PSPM18,PSPM19,SPP19b,MSPPHP19,MSHPP19} and target control with instantaneous perturbations~\citep{BPSP19} for asynchronous Boolean networks.
In view of the difficulties and expenses in conducting biological experiments, 
our methods compute the minimal control sets, which can be easily translated for wet-lab validation.

Cells in tissues and in culture normally exist as a population of cells, 
corresponding to different stable steady states~\citep{SC14}. 
There is a need of target control methods to compute a subset of nodes, 
the control of which can always drive the system from any initial state to a desired target attractor.  
The target control method developed in our previous work~\citep{BPSP19} adopts instantaneous perturbations, 
that are only applied instantaneously, but at a cost, 
rather larger number of control nodes are required than temporary and permanent perturbations~\citep{SPP19b}. 
Moreover, it is difficult to guarantee that all the perturbations take effect at the same time in biological experiments.   
Thus, target control with temporary perturbations is more appealing.

In this paper, we develop a target control method with temporary perturbations for asynchronous Boolean networks. 
Our idea is to find a control $C=(\zero, \one)$, which is a tuple of two sets,  
such that the application of $C$ -- setting the value of a node, 
whose index is in $\zero$ (or $\one$), to $0$ (or $1$) -- 
can drive the network from any initial state $s$ in the state space $S$ 
to an intermediate state $s'$ in the weak basin of the target attractor. 
We hold the control $C$ for sufficient time and let the network evolve to a state in the strong basin of the target attractor.
After that, the control can be released and the network will eventually and surely reach the target attractor. 
Since the network can take any state $s\in S$ as an initial state, 
the possible intermediate states form a subset $S'$ of $S$, called {\it schema}. 
According to our previous work~\citep{SPP19b}, we know that 
all the intermediate states should fall into the weak basin of the target attractor. 
Therefore, we partition the weak basin into a set of mutually disjoint schemata. 
Each schema results in a candidate control, which is further minimised and verified. 
Clinical applications are highly time-sensitive, controlling more nodes may shorten the period of time for generating sufficient desired cells~\citep{GD19}. 
Hence, we integrate our method with a threshold $\zeta$ on the number of perturbations.  
By increasing $\zeta$, we can obtain solutions with at most $\zeta$ perturbations.   
It is worth noting that more perturbations may cause a significant increase in the experimental cost, hence, 
the parameter $\zeta$ should be considered individually based on specific experimental settings.

We have implemented our method and compared its performance with 
the stable motif-based control (SMC)~\citep{ZA15} on various real-life biological networks, 
as both methods focus on temporary target control of asynchronous Boolean networks.   
The results show that our method outperforms SMC in terms of the computational time for most of the networks. 
Both methods find a number of valid temporary controls, 
but our method is able to identify more controls with fewer perturbations for some networks. 
Another interesting observation is that the number of required perturbations is often quite small compared to the sizes of the networks.  
This agrees with the empirical findings that the control of few nodes can reprogram biological networks~\citep{MS11}. 

\section{Background and Notations}\label{sec:background}
In this section, we give preliminary notions of Boolean networks. 
Let $[n]$ denote the set of positive integers $\{1,2, \ldots, n\}$. 

\subsection{Boolean networks}
\label{ssec:bn}
A Boolean network (BN) describes elements of a dynamical system with binary-valued nodes and interactions between elements with Boolean functions. It is formally defined as:

\begin{definition}[Boolean networks]
A Boolean network is a tuple $G = (X,F)$ where $X=\{x_1,x_2,\ldots, x_n\}$,
such that $x_i, i\in[n]$ is a Boolean variable and $F=\{f_1,f_2,\ldots,f_n\}$ is a set of Boolean functions over $X$.
\end{definition}

A Boolean network $G = (X,F)$ can be viewed as a directed graph $\mathcal{G} = (V,\mathcal{E})$, 
called the {\em dependency graph} of $G$,
where $V=\{v_1,v_2\ldots, v_n\}$ is the set of {\em nodes}. 
Node $v_i \in V$ corresponds to variable $x_i \in X$. 
For every $i,j\in [n]$, there is a directed edge from $v_j$ to $v_i$ if and only if
$f_i$ depends on $x_j$.  
For the rest of the exposition, we assume an arbitrary but fixed network $G=(X,F)$ of $n$ variables is given to us.  
For all occurrences of $x_i$ and $f_i$, we assume $x_i$ and $f_i$ are elements of $X$ and $F$, respectively.
A {\em state} $s$ of $G$ is an element in $\{0,1\}^n$.
Let $ S$ be the set of states of $G$. 
For any state $s=(s[1],s[2],\ldots,s[n])$, and for every $i\in[n]$, the value of $s[i]$,
represents the value that $x_i$ takes when the network is in state $s$.
For some $i\in[n]$, suppose $f_i$ depends on $x_{i_1},x_{i_2},\ldots, x_{i_k}$. 
Then $f_i(s)$ will denote the value $f_i(s[i_1],s[i_2],\ldots, s[i_k])$  
and $x_{i_1},x_{i_2},\ldots, x_{i_k}$ are called {\it parent nodes} of $x_i$. 
For two states $s,s'\in S$, the {\em Hamming distance} between $s$ and $s'$ is denoted as $\hd(s,s')$. 
\begin{definition}[Control] \label{def:control}
A control $C$ is a tuple $(\zero,\one)$, where $\zero, \one \subseteq [n]$ and $\zero$ and $\one$ are mutually disjoint (possibly empty) sets of indices of nodes of a Boolean network $G$. 
The size of the control $C$ is defined as $|C|=|\zero|+|\one|$. 
Given a state $s\in S$, the application of $C$ to $s$, denoted as $C(s)$, 
is defined as a state $s'\in S$, such that $s'[i]=0=1-s[i]$ for $i \in \zero$ 
and $s'[i]=1=1-s[i]$ for $i \in \one$. 
$s'$ is called the intermediate state w.r.t. $C$.
\end{definition}

The control can be lifted to a subset of states $S' \subseteq S$. 
Given a control $C=(\zero,\one)$, $C(S')=S''$, where $S''=\{s''\in S|s''=C(s'), s'\in S'\}$. 
$S''$ includes all the intermediate states with respect to $C$. 
The application of $C$ results in a new Boolean network, defined as follows.

\begin{definition}[Boolean networks under control] \label{def:BNcontrol}
Let $C=(\zero,\one)$ be a control and $G=(X,F)$ be a Boolean network. 
The Boolean network $G$ under control $C$, denoted as $G|_C$, is defined as a tuple $G|_C=(\hat{X},\hat{F})$, 
where $\hat{X}=\{\hat{x}_1,\hat{x}_2,\ldots, \hat{x}_n\}$ and $\hat{F}=\{\hat{f}_1,\hat{f}_2,\ldots,\hat{f}_n\}$, 
such that for all $i \in [n]$: \\
(1) $\hat{x}_i=0$ if $i\in \zero$, $\hat{x}_i=1$ if $i\in \one$, and $\hat{x}_i=x_i$ otherwise; \\
(2) $\hat{f}_i=0$ if $i\in \zero$, $\hat{f}_i=1$ if $i\in \one$, and $\hat{f}_i=f_i$ otherwise.
\end{definition}

The state space of $G|_C$, denoted $ S|_C$,  is derived by fixing the values of the variables in the set $C$ to their respective
values and is defined as $ S|_C=\{ s\in S\ |\  s[i]=1 \text{ if } i\in \one \text{ and }  s[j]=0 \text{ if } j\in \zero\}$. 
Note that $ S|_C\subseteq  S$.
For any subset $S'$ of $S$, we let $ S'|_C = S'\cap S|_C$. 

\begin{figure}[!t]
\begin{minipage}[b]{0.6\linewidth}
\centering
\begin{tabular}{l}
 $f_1=x_2$ \\
 $f_2=x_1$ \\
 $f_3=x_2 \land x_3$\\
\end{tabular}
\\(a) \\[3pt]
\end{minipage}
\begin{minipage}[b]{0.35\linewidth}
\centering
\scalebox{.95}{
\begin{tikzpicture}
\tikzset{node distance=1.2cm and 1.2cm}
\node        (s1)                  {$~x_1~$};
\node        (s2)  [right=of s1]   {$~x_2~$};
\node        (s3)  [right=of s2]   {$~x_3~$};

 \path     
 (s3)  edge [<-,loop above] node {} (s3)
 (s1)  edge [->,bend left] node {} (s2)
 (s2)  edge [->,bend left] node {} (s1)
 (s2)  edge [->] node {} (s3)
 ;
\end{tikzpicture}}
\\ (b)
\end{minipage}
\begin{minipage}{0.6\linewidth}
\centering
\includegraphics[width=.9\textwidth]{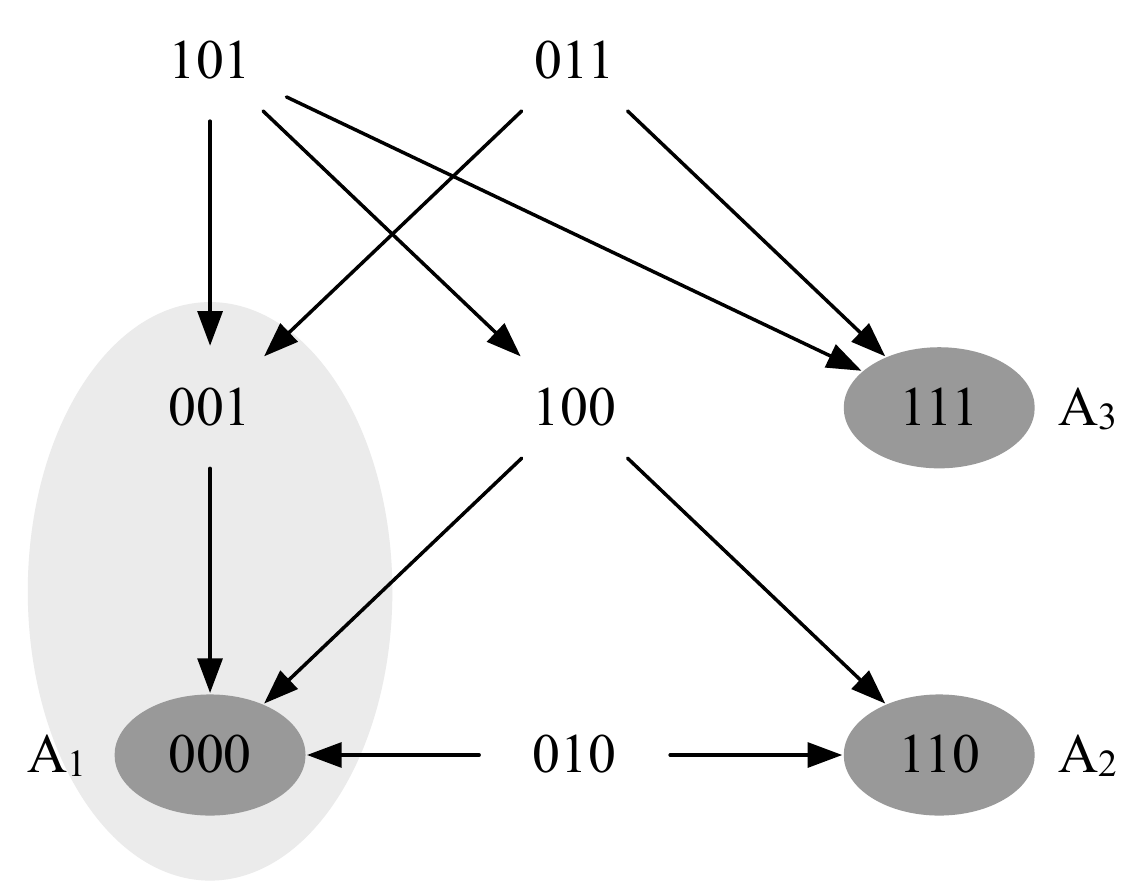}
\\ (c)
\end{minipage}
\begin{minipage}{0.35\linewidth}
\centering
\includegraphics[width=.8\textwidth]{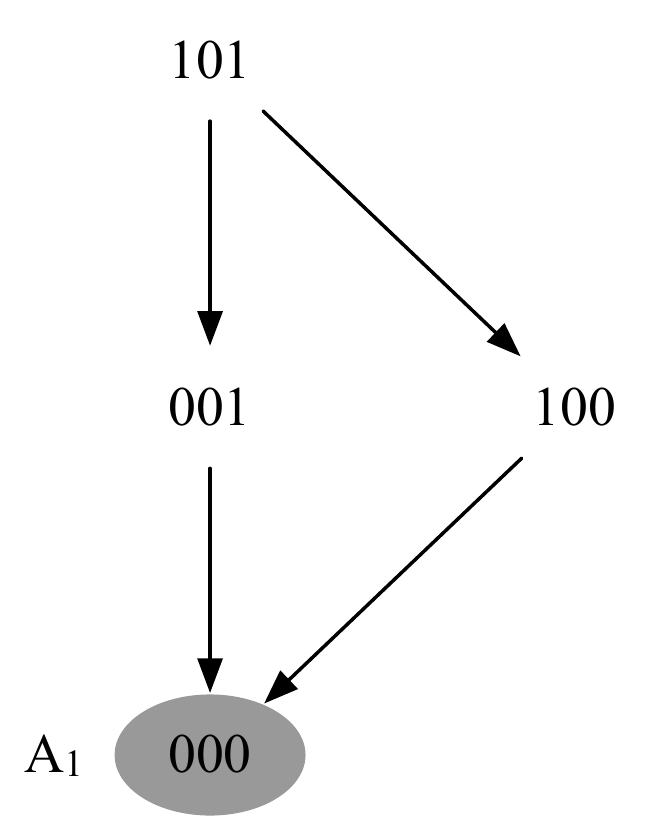}
\\ (d)
\end{minipage}
\caption{(a) Boolean functions, (b) the dependency graph, (c) transition system $\ts$ and 
(d) transition system under control $\ts|_C$ for Example~\ref{eg:bn}. We omit selfloops for all the states except for state $101$ in $(c)$.}
\label{fig:bn}
\end{figure}
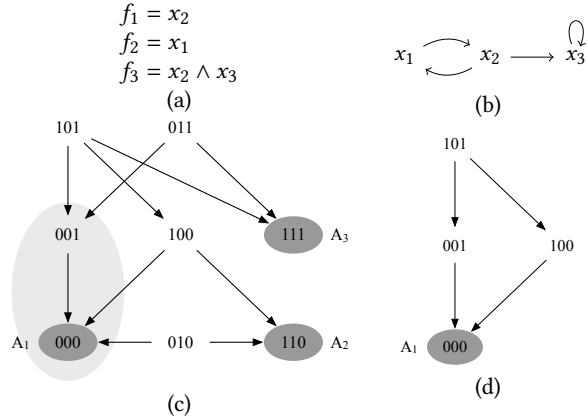

\subsection{Dynamics of Boolean networks}
\label{ssec:dynamics}
In this section and the next section, 
we define several notions that can be interpreted on both $G$ and $G|_C$. 
We use the generic notion $G=(X,F)$ to represent either $G=(X,F)$ or $G|_C=(\hat{X},\hat{F})$. 
We assume that a Boolean network $G=(X,F)$ evolves in discrete time steps.
It starts in an initial state $s_0$ and its state changes in every time step according to the update functions $F$.
Different updating schemes lead to different dynamics of the network~\citep{MPSY18,ZH14}.
In this work, we are interested in the {\em asynchronous updating scheme} as it allows biological processes to happen at different classes of time scales and thus is more realistic.

The asynchronous evolution of $G$ is a function $\update: \naturals \rightarrow \wp( S)$
such that $\update(0)=\{s_0\}$ and for every $j\geq 0$,
if $s \in \update(j)$ then $s'\in \update(j+1)$ is a possible {\em next state} of $s$
iff either $\hd(s,s') = 1$ and $s'[i]=f_i(s)=1-s[i]$  
or $\hd(s,s')=0$ and there exists $i$ such that $s'[i]=f_i(s)=s[i]$.
It is worth noting that the asynchronous dynamics is non-deterministic.  
At each time step, only one node is randomly selected to update its value 
and a different choice may lead to a different next state $s'\in\update(j+1)$. 
Henceforth, when we talk about the dynamics of $G$, we shall mean the asynchronous dynamics. 
The dynamics of a Boolean network can be described as a {\em transition system (TS)}.
\begin{definition}[Transition system of Boolean networks]\label{def:ts}
The transition system of a Boolean network $G$, denoted as $\ts$, is a tuple $(S,E)$,
where the vertices are the set of states $ S$ and for any two states $s$ and $s'$
there is a directed edge from $s$ to $s'$, denoted $s \rightarrow s'$
iff $s'$ is a possible next state of $s$ according to the asynchronous evolution function $\update$ of $ G$.
\end{definition}
Similarly, we denote the transition system of a Boolean network under control, $G|_C$, as $\ts|_C$. 

\begin{example} \label{eg:bn}
Consider a network $G=(X,F)$, where $X=\{x_1,x_2,x_3\}$, $F=\{f_1,f_2,f_3\}$,
and $f_1=x_2$,
$f_2=x_1$ and
$f_3=x_2 \land x_3$. 
The dependency graph of the network $\mathcal{G}$ and its associated transition system $\ts$ are given in Fig.~\ref{fig:bn} $(b)$ and $(c)$. 
Given a control $C=(\zero, \one), \zero=\{2\}, \one=\emptyset$ (i.e., $\{x_2=0\}$), 
the transition system under control $\ts|_C$ is given in Fig.~\ref{fig:bn} $(d)$. 
\end{example}

\subsection{Attractors and basins}
\label{ssec:attractor}
A {\em path} $\path$ from a state $ s$ to a state $ s'$ is a (possibly empty) sequence of transitions from $ s$ to $ s'$ in $\ts$, 
denoted $\path=s \rightarrow s_1 \rightarrow \ldots \rightarrow s'$.
A path from a state $ s$ to a subset $ S'$ of $ S$ is a path from $ s$ to any state $ s'\in  S'$. 
An {\em infinite path} $\path$ from $s$, $\path=s \rightarrow s_1 \rightarrow \ldots $, 
is a sequence of infinite transitions from $s$.   
A state $s' \in S $ appears {\it infinitely often} in $\path$ if for any $i \geq 0$, 
there exists $j \geq i$ such that $s_j=s'$. 
We assume every infinite path $\path$ is {\it fair} -- 
for any state $s'$ that appears infinitely often in $\path$, 
every possible next state $s''$ of $s'$ also appears infinitely often in $\path$. 
For a state $ s\in S$, $\reach(s)$ denotes the set of states $ s'$ such that there is a path 
from $s$ to $s'$ in $\ts$. 

\begin{definition}[Attractor]\label{def:attractor}
An attractor $A$ of $\ts$ (or of $G$) is a minimal non-empty subset of states of $ S$ such that for every state $ s\in A,~\reach(s)=A$.
\end{definition}

\begin{figure}
\centering
\includegraphics[width=0.48\textwidth]{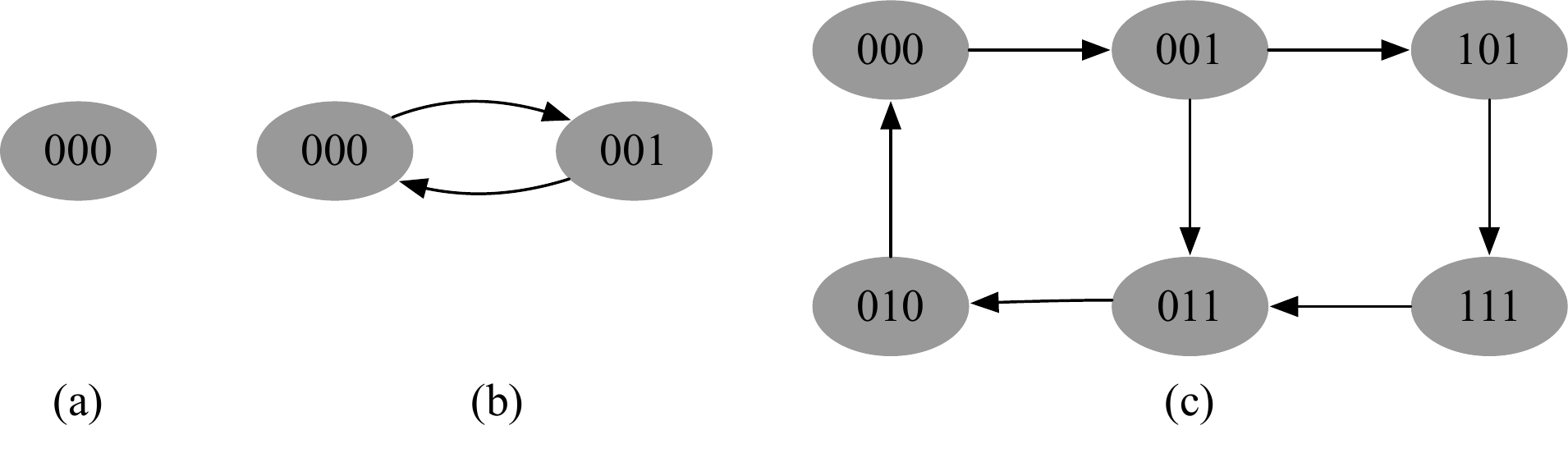} 
\caption{Different types of attractors of an asynchronous Boolean network. We omit selfloops for all the states.}
\label{fig:attractors}
\end{figure}

Attractors are hypothesised to characterise the steady-state behaviour of the network. 
Any state which is not part of an attractor is a transient state.
An attractor $A$ of $\ts$ is said to be reachable from a state $ s$ 
if $\reach(s)\cap A\neq\emptyset$.
The network starting at any initial state $s_0 \in S$ will eventually end up in one of the attractors of $\ts$ and remain there forever unless perturbed.
Under asynchronous updating scheme, there are singleton attractors and cyclic attractors.  
Cyclic attractors can be further classified into:  
(1) a simple loop, in which all the states form a loop and every state appears only once per traversal through the loop; 
and (2) a complex loop, which has intricate topology and includes several loops. 
Fig.~\ref{fig:attractors} $(a)$, $(b)$ and $(c)$  
show a singleton attractor, a simple loop and a complex loop, respectively. 
Let $\mathcal{A}$ denote all the attractors of $\ts$. 
For an attractor $A,~A\in \mathcal{A}$, 
we define its {\it weak basin} as $\bas^W_\ts(A) = \{s\in S\ |\ \reach(s)\cap A\neq \emptyset\}$; 
the {\it strong basin} of $A$ is defined as $\bas^S_\ts(A) = \{s\in S\ |\ \reach(s)\cap A\neq \emptyset 
\text{ and } \reach(s) \cap A' = \emptyset  \text{ for any } A' \in \mathcal{A}, A'\neq A\}$. 
Intuitively, the weak basin of $A$, $\bas^W_\ts(A)$, 
contains all the states $s$ from which there exists at least one path to $A$,  
and there may also exist paths from $s$ to other attractor $A'~(A' \neq A)$ of $\ts$. 
The strong basin of $A$, $\bas^S_\ts(A)$, consists of all the states from which there only exist paths to $A$. 

\begin{example} \label{eg:bn2}
The network in Example~\ref{eg:bn} has three attractors $A_1=\{000\}$, $A_2=\{110\}$ and $A_3=\{111\}$, 
indicated as dark grey nodes in Fig.~\ref{fig:bn}$(c)$. 
For attractor $A_1$, its strong basin $\bas^S_\ts(A_1)=\{000, 001\}$ is shown as the shaded grey region; 
its weak basin contains six states, i.e. $\bas^W_\ts(A_1)=\{000, 001, 101, 011, 100, 010\}$.
We can see that only attractor $A_1$ is preserved in $\ts|_C$ in Fig.~\ref{fig:bn}$(d)$.  
\end{example}

\subsection{The control problem}
\label{ssec:problem}
As described in the introduction, many devastating diseases, such as Parkinson's disease and Alzheimer's disease, 
are caused by a deficiency of particular cells. 
Cell reprogramming can transform abounding somatic cells into the desired cell type. 
In the context of Boolean networks, this process is, indeed, 
stirring the dynamics of the network from a source attractor to a desired target attractor. 
However, cells in culture and in situ are usually not isolated but exist in a population consisting of various cell phenotypes or even transient cell states. 
Hence, it is important to develop a target control method to identify key nodes 
that can guide the network towards a desired target attractor from any other distinct steady states or transient states. 

This can be defined as a {\it target control} problem:  
given a Boolean network $G$ and a target attractor $A_t$, finding a control $C$, 
the application of which can drive the network from any source state $s \in S$ to $A_t$. 
When the source state $s$ is fixed, finding a control $C$ to drive the network from $s$ to $A_t$, 
is a {\it source-target control} problem. 
Based on the application time of control, we have:
(1) {\it temporary control} - perturbations are applied for a finite (possibly zero) number of steps and then released;
(2) {\it permanent control} - perturbations are applied for all the following steps. 
When perturbations are applied instantaneously, we call it {\it instantaneous control}, 
which is a special case of temporary control. 
Temporary control has shown its apparent advantages in reducing the number of perturbations~\citep{SPP19b}, 
thus in this work, we focus on {\it temporary target control}, formally defined as follows.

\begin{definition}[Temporary target control]\label{def:ttc}
A temporary target control is a control $C=(\zero, \one)$, such that there exists a $t_0>0$, for all $t>t_0$, the network always reaches the target attractor $A_t$ on the application of $C$ to any source state $s\in S$ for $t$ steps.
\end{definition}

\section{Results}\label{sec:results}
\begin{algorithm*}[!t]
\centering
\begin{algorithmic}[1]
\Procedure{{\sc Temp\_Target\_Control}}{$G, A_t$}
	\State initialise $\mathcal{L}:=\emptyset$ and $\Omega := \emptyset$ to store valid temporary control sets and the checked control sets, respectively.
	\State \label{line:input} $I, I^\mathit{ns} :=${\sc Comp\_input\_nodes}$(G)$ \hfill {\it \footnotesize {//compute input nodes $I$ and non-specified input nodes $I^\mathit{ns}$}.}
	\State \label{line:sb} $\sbbas:=${\sc Comp\_SB}$(F,A_t)$ \hfill{\it //strong basin of $A_t$ in $\ts$}
	\State\label{line:wb} $\wb:=${\sc Comp\_WB}$(F,A_t)$ \hfill{\it //weak basin of $A_t$ in $\ts$}
	\State \label{line:can}  $\mathcal{W}:=${\sc Comp\_schemata}$(\wb)$, $m:= |\mathcal{W}|$ 
	\State generate a vector $\Theta$ of length $m$ and set all the elements to $\false$
	\hfill {\it $\Theta[i]$ indicates if $W_i$ can be skipped or not.}
	\State $\zeta:=n$ \hfill{\it set an initial threshold on the number of perturbations. $n$ is the size of the network.}

	\For {$i = 1: m$} \hfill {\it // traverse the schemata }
		\State \label{line:skip} {\sf if} {$\Theta[i]=\true$}, {\sf then} {\sf continue}
		\State \label{line:sup} $C_i:=${\sc Comp\_support\_variables}$(W_i)$ \hfill{\it // $C_i:=(\zero_i, \one_i)$}
			\State \label{line:ec} $C^e_i := (\zero_i \cap I^\mathit{ns}, \one_i \cap I^\mathit{ns})$, 
			$C^r_i := (\zero_i \setminus I, \one_i \setminus I)$ 
			\hfill{\it //essential control nodes and non-input nodes in $C_i$}
			\State $k:=0$, $\mathit{isValid} := \false$
			\While{$\mathit{isValid} = \false$ and $k \leq \min(\zeta -|C^e_i|, |C^r_i|)$}
				\State $\mathcal{C}^\mathit{sub}_i := ${\sc Comp\_subsets}$(C^r_i,k)$  \hfill {\it //compute subsets of $C^r_i$ of size $k$.}
				\For {$C^\mathit{sub}_j \in \mathcal{C}^\mathit{sub}_i$}
					\State $C_i^j := C^\mathit{sub}_j \cup C^e_i$, $\Phi :=C_i^j(S) $ \hfill {\it // $\Phi$ represents the intermediate states w.r.t. $C_i^j$.}
					\If {$C_i^j \notin \Omega$}  \hfill{\it //  $C_i$ has not been checked.}
						\State $\mathit{isValid} := ${\sc Verify\_TTC}$(F, C_i^j,\sbbas, \Phi)$
						\State add $C_i^j$ to $\Omega$. 
						\If{$\mathit{isValid}=\true$}
							\State add $C^j_i$ to $\mathcal{L}$, $\zeta := \min (\zeta, |C^j_i|)$
							\State \label{line:remainW} $\Theta[z]:=\true$ if $W_z \subseteq \Phi$ for $z\in [i+1,m]$
							\hfill {\it // if a schema $W_z$ is a subset of $\Phi$, it will be skipped.}
						\EndIf
					\EndIf
				\EndFor
				\State {\sf if} $\mathit{isValid} = \false$, {\sf then} $k:=k+1$
			\EndWhile
	\EndFor
	\State \Return $\mathcal{L}$
\EndProcedure
\end{algorithmic}
\caption{Temporary Target Control}
\label{alg:ttc}
\end{algorithm*}

\begin{algorithm}[t]
\centering
\begin{algorithmic}[1]
\Procedure{{\sc Verify\_TTC}}{$F, C, \sbbas, \Phi$} \label{line:verify}
	\State $\mathit{isValid} := \false$
	\If {$\Phi \subseteq \sbbas$}
		\State $\mathit{isValid} = \true$
	\Else
		\State $\sbbas|_C :=${\sc Comp\_state\_control}$(C, \sbbas)$ \hfill {\it //compute the remaining strong basin w.r.t. $C$ in $\ts|_C$}
		\State $F|_C := ${\sc Comp\_Fn\_control}$(C, F)$ 
		\State $\bas^S_{\ts|_C}(SB|_C) := ${\sc Comp\_SB}$(F|_C, \sbbas|_C)$ 
		\If{$\Phi \subseteq \bas^S_{\ts|_C}(SB|_C)$}
			\State $\mathit{isValid} = \true$
		\EndIf
	\EndIf
	\State \Return $\mathit{isValid}$
\EndProcedure
\end{algorithmic}
\caption{Verification of Temporary Target Control}
\label{alg:verify}
\end{algorithm}

In this section, we shall develop a method to solve the temporary target control problem. 
First, we introduce the following lemma, which is crucial for the development of the method.    
\begin{lemma}\label{prop:tc}
A control $C=(\zero, \one)$ is a temporary target control to a target attractor $A_t$ from any source state $s\in S$ iff 
$\bas^S_\ts(A_t) \cap S|_C \neq \emptyset$ and $C(S) \subseteq \bas^S_{\ts|_C}(\bas^S_\ts(A_t) \cap S|_C)$.
\end{lemma}

Instead of presenting a formal proof for Lemma~\ref{prop:tc},
we give an intuitive explanation below.  
Definition~\ref{def:BNcontrol} shows that the application of a control $C$ results in 
a new Boolean network $G|_C$ and the state space is restricted to $S|_C$. 
To guarantee the inevitable reachability of $A_t$, by the time we release the control, 
the network has to reach a state $s$ in the strong basin of $A_t$ 
w.r.t. the original transition system $\ts$, i.e. $\bas^S_\ts(A_t)$, 
from which there only exist paths to $A_t$. 
This requires the remaining strong basin in $S|_C$, i.e. $(\bas^S_\ts(A_t) \cap S|_C)$, is a non-empty set; 
otherwise, it is not guaranteed to reach $A_t$. 
Furthermore, the condition $C(S) \subseteq \bas^S_{\ts|_C}(\bas^S_\ts(A_t) \cap S|_C)$ ensures 
any possible intermediate state $s' \in C(S)$ is in the strong basin of the remaining strong basin $(\bas^S_\ts(A_t) \cap S|_C)$ in the transition system under control $\ts|_C$, 
so that the network will always evolve to the remaining strong basin.  
Once the network reaches the remaining strong basin, 
the control can be released and the network will evolve spontaneously towards the target attractor $A_t$. 
Based on the definition of the weak basin, 
it is sufficient to search the weak basin $\bas^W_\ts(A_t)$ for temporary target control.

A noteworthy point is that temporary control needs to be released once the network reaches a state in $(\bas^S_\ts(A_t) \cap S|_C)$. 
On one hand, Lemma~\ref{prop:tc} guarantees that partial strong basin of $A_t$ in $\ts$ is preserved in $\ts|_C$, while it does not guarantee the presence of $A_t$ in $\ts|_C$. 
In that case, the control $C$ has to be released at one point to recover the original $\ts$, which at the same time retrieves $A_t$.  
On the other hand, in clinic, it is preferable to eliminate human interventions to avoid unforeseen consequences.
Concerning the timing to release the control, 
since it is hard to interpret theoretical time steps in diverse biological experiments, 
it would be more feasible for biologists to estimate the timing based on empirical knowledge and specific experimental settings. 

Previously, we have developed efficient decomposition-based algorithms to compute the exact basins of an attractor, which exploit both the structural and dynamical properties of the network~\citep{PSPM18,PSPM19}.   
In the algorithm we develop here, we shall use these procedures to compute the weak basin and the strong basin of an attractor and refer them as {\sc Comp$\_$WB} and {\sc Comp$\_$SB}, respectively. 
Next, we define the {\it projection} of a state $s\in S$ to a subset $B$ of $[n]$, 
which represents the indices of a subset of nodes $X' \subseteq X$ as follows.

\begin{definition}[Projection]
Let $X'=\{x_{i_1}, x_{i_2}, \ldots, x_{i_k}\}$ be a subset of $X$ 
and $B=\{i_1, i_2, \ldots, i_k\}$ be the set of indices of $X'$. 
The projection of a state $s$ to $B$, is an element of $\{0,1\}^k$, 
defined as $s|_{B}=(s[i_1],s[i_2], \ldots, s[i_k])$. 
The projection is lifted to a subset $S'$ of $S$ as $S'|_{B}=\{s|_{B}|s \in S'\}$.
\end{definition}
Given a control $C=(\zero, \one)$, the possible intermediate states with respect to $C$, denoted $S'=C(S)$, 
form a {\it schema}, and can be defined as follows.

\begin{definition}[Schema]
A subset $S'$ of $S$ is a schema if there exists a triple $M=(\zero, \one, \D)$, 
where $\zero \cup \one \cup \D = [n]$, 
$\zero, \one$ and $\D$ are mutually disjoint (possibly empty) set of indices of nodes of $G$, 
such that $S'|_\zero=\{0\}^{|\zero|}$, $S'|_\one=\{1\}^{|\one|}$ and $S'|_\D=\{0,1\}^{|\D|}$. 
$\zero, \one$ and $\D$ are called off-set, on-set and don't-care-set of $S'$, respectively.  
The elements in $\zero \cup \one$ are called indices of support variables of $S'$. 
\end{definition}

Intuitively, for any node $x_i, i \in \zero$, it has a value of $0$ in any state $s\in S'$; 
for any node $x_i, i \in \one$, it has a value of $1$ in any state $s\in S'$. 
The projection of $S'$ to the don't-care-set $\D$ contains all combinations of binary strings with $|\D|$ bits.
Thus, any schema $S'$ is of size $2^{|\D|}$. 
Since the total number of nodes $n = |\zero| + |\one| + |\D|$ is fixed, 
a larger schema implies more elements in $\D$ and fewer elements in $\zero \cup \one$.

\begin{example}\label{eg:bn_set} 
To continue with Example~\ref{eg:bn2}, the set $W_1=\{000,001,010,011\}$ is a subset of the weak basin of $A_1$ in $\ts$. 
There exists a triple $M_1=(\zero_1, \one_1, \D_1)$, 
where $\zero_1=\{1\}$, $\one_2=\emptyset$ and $\D_2=\{2,3\}$, 
such that $W_1|_{\zero_1}=\{0\}$,  $W_1|_{\one_1}=\emptyset$ and $W_1|_{\D_1}=\{00,01,10,11\}$. 
Therefore, $W_1$ is a schema. 
Let us denote the value of $x_i$, i in $\zero_1$, $\one_1$ and $\D_1$, as $0, 1$ and $*$, respectively. 
Then, $W_1$ can be represented as $0**$. 
\end{example}

The notion of schema leads the way to find temporary target control. 
Each schema $W_i$ of the weak basin $\bas^W_\ts(A_t)$ gives a candidate temporary target control $C_i=(\zero_i, \one_i)$ 
for further optimisation and validation. 
A larger schema results in a smaller control set. 
To explore the entire weak basin $\bas^W_\ts(A_t)$, 
we partition it into a set of mutually disjoint schemata 
$\mathcal{W}=\{W_1, W_2, \ldots, W_m\}$, $W_1 \cup W_2 \cup \ldots \cup W_m = \bas^W_\ts(A_t)$. 
Each $W_i, i \in m$ is one of the largest schemata in $\bas^W_\ts(A_t) \setminus (W_1 \cup \ldots \cup W_{i-1})$.
For $W_i$, the indices of its support variables in $\zero_i$ and $\one_i$ form a candidate control $C_i=(\zero_i,\one_i)$.  
Each candidate control $C_i$ is primarily optimised based on the properties of input nodes. 
Because input nodes do not have any predecessors,  
it is reasonable to assume that specified input nodes $I^s$ are redundant control nodes,  
while non-specified input nodes $I^\mathit{ns}$ are essential for control. 
For the remaining non-input nodes in $C_i$, denoted $C^r_i$, 
we verify its subsets of size $k$ based on Lemma~\ref{prop:tc} from $k=0$ with an increment of $1$, 
until we find a valid solution.

To further improve the efficiency of our method, we use binary decision diagram (BDD) as a symbolic representation of large state space.  
The size of a BDD is determined by both the set of states  
being represented and the chosen ordering of the variables.
In BDD, a schema is represented as a {\it cube} and each state is the smallest cube, also called a {\it minterm}. 
To compute the largest schema $S_i$ of $S$ is equivalent to compute the largest cube of $S$. 
The partitioning of the weak basin into schemata is then transformed into a cube cover problem in BDD. 
A different variable ordering may lead to a different partitioning. 
Given a fixed ordering, the partitioning remains the same. 
Although finding the best variable ordering is NP-hard, 
there exist efficient heuristics to find the optimal ordering.
In this work, we compute a partitioning under one variable ordering as provided by the CUDD package~\citep{cudd}
and compute the smallest subsets of candidate controls that are valid temporary target control sets.

Algorithm~\ref{alg:ttc} implements the idea in pseudo-code. 
It takes as inputs the Boolean network $G=(X, F)$ and the target attractor $A_t$.  
It first initialises two vectors $\mathcal{L}$ and $\Omega$ to store valid controls and the checked controls, respectively. 
(We use $\Omega$ to avoid duplicate control validations.)
Then, it computes input nodes $I$ and the non-specified input nodes $I^\mathit{ns},~ 
I^\mathit{ns} \subseteq I$ (line~\ref{line:input}). 
The strong basin $\sbbas$ and the weak basin $\wb$ of $A_t$ of $\ts$ are computed using the decomposition-based procedures
{\sc Comp\_SB} and {\sc Comp\_WB} developed in~\citep{PSPM18,PSPM19} (lines~\ref{line:sb}-\ref{line:wb}). 
The weak basin $\wb$ is then partitioned into $m$ mutually disjoint schemata 
with procedure {\sc Comp\_schemata}. 
Realisation of this procedure relies on the function 
to compute the largest cube provided by the CUDD package~\citep{cudd}. 
For each schema $W_i$, 
the indices of its support variables computed by procedure {\sc Comp\_support\_variables} 
form a candidate control $C_i$ (line~\ref{line:sup}). 
The essential control nodes $C^e_i$ of $C_i$ consist of the non-specified input nodes 
and the non-input nodes in $C_i$ constitute a set $C^r_i$ for further optimisation (line $12$). 
We search for the minimal subsets of $C^r_i$ starting from size $k=0$ with an increment of $1$ 
and verify whether the union of a subset $C_j^\mathit{sub}$ of $C_i^r$
and the essential nodes $C^e_i$, namely $C_i^j = C_j^\mathit{sub} \cup C_i^e$, 
is a valid temporary target control using procedure {\sc Verify\_TTC} in Algorithm~\ref{alg:verify}.
If $C_i^j$ is valid, save it to $\mathcal{L}$.  
When all the subsets have been traversed or a valid control has been found, 
we proceed to the next schema $W_{i+1}$. 
In the end, all the verified temporary target controls are returned.

The most time-consuming part of our method lies in the verification process. 
As shown in Algorithm~\ref{alg:verify}, for each candidate control $C$, 
we need to reconstruct the associated transition relations $F|_C$ and compute the strong basin of the remaining strong basin in $\ts|_C$, i.e. $\bas^S_{\ts|_C}(SB|_C)$ (lines $6$ and $7$ of Algorithm~\ref{alg:verify}). 
Even though we have developed an efficient method for basin computation, 
the computational time of Algorithm~\ref{alg:verify} still increases when the network size grows. 
To improve the efficiency, we propose two heuristics: 
(1) skip a schema $W_z$ (line $10$ and $23$ of Algorithm~\ref{alg:ttc})
if it is a subset of intermediate states $\Phi$ of a pre-validated control $C_i^j$ (line $23$ of Algorithm~\ref{alg:ttc});   
and (2) set a threshold $\zeta$ on the number of perturbations,  
keep $\zeta$ updated with the smallest size of valid temporary target control $C_i^j$ (line $22$ of Algorithm~\ref{alg:ttc}) 
and only compute control sets with at most $\zeta$ perturbations.

Algorithm~\ref{alg:ttc} is easily adapted to solve target control problem with instantaneous perturbations
by focusing on the schemata of the strong basin of $A_t$. 
In this way, we don't need to use Algorithm~\ref{alg:verify} for additional verification
and the indices of support variables of each schema form an instantaneous control.

\section{Evaluation}\label{sec:evaluation}
\begin{table*}[!t]
\centering
\begin{tabular}{|L{2cm}|R{0.7cm}R{0.7cm}|R{0.9cm}R{0.6cm}|R{0.9cm}R{0.6cm}|R{1.1cm}R{1.1cm}|R{1cm}R{1cm}|}
\hline
 \multicolumn{1}{|l|}{\multirow{3}{*}{Network}} 
& \multicolumn{1}{r}{\multirow{3}{*}{$\#$nodes}}& \multicolumn{1}{r|}{\multirow{3}{*}{$\#$edges}}  
& \multicolumn{4}{c|}{Number of attractors} &  \multicolumn{4}{c|}{Time (seconds)}\\  \cline{4-11}
&&& \multicolumn{2}{c|}{TTC} &  \multicolumn{2}{c|}{SMC}& \multicolumn{2}{c|}{Attractor detection} &  \multicolumn{2}{c|}{Control} 
\\ \cline{4-11}
&&& \multicolumn{1}{c|}{\multirow{1}{*}{$\#$singleton}}	&\multicolumn{1}{c|}{\multirow{1}{*}{$\#$cyclic}}	 
& \multicolumn{1}{c|}{\multirow{1}{*}{$\#$singleton}}	&\multicolumn{1}{c|}{\multirow{1}{*}{$\#$quasi}}
& \multicolumn{1}{c|}{\multirow{1}{*}{TTC}}	&\multicolumn{1}{c|}{\multirow{1}{*}{SMC}}
& \multicolumn{1}{c|}{\multirow{1}{*}{TTC}}	&\multicolumn{1}{c|}{\multirow{1}{*}{SMC}}

\\ \hline
myeloid	&	$11$	&	$30$	&	$6$	&	$0$	&	$6$	&	$0$	&	$0.002$	&	$7.100$	&	$0.025$	&	$7.710$ \\ 
apoptosis	&	$12$	&	$26$	&	$2$	&	${\bf 1}$	&	$2$	&	${\bf 1}$	&	$0.004$	&	$2.423$	&	$0.010$	&	$2.679$ \\ 
cardiac	&	$15$	&	$39$	&	$6$	&	$0$	&	$6$	&	$0$	&	$0.004$	&	$10.710$	&	$0.200$	&	$10.279$ \\ 
ERBB	&	$20$	&	$52$	&	$3$	&	$0$	&	$3$	&	$0$	&	$0.004$	&	$6.400$	&	$0.105$	&	$5.788$ \\ 
HSPC-MSC&	$26$	&	$81$	&	$2$	&	$2$	&	$2$	&	$2$	&	$0.101$	&	$33.910$	&	$0.099$	&	$11.433$ \\ 
PC12	&	$33$	&	$62$	&	$7$	&	$0$	&	$7$	&	$0$	&	$0.013$	&	$84.904$	&	$14.953$	&	$191.299$ \\ 
hematopoiesis	&	$33$	&	$88$	&	$5$	&	$0$	&	-	&	-	&	$0.452$	&	-	&	$97.773$	&	- \\ 
bladder	&	$35$	&	$116$	&	$3$	&	$1$	&	$3$	&	$1$	&	$0.735$	&	$25.662$	&	$2.181$	&	$34.035$ \\ 
MAPK	&	$53$	&	$105$	&	$2$	&	$0$	&	$2$	&	$0$	&	$1.749$	&	$6.461$	&	$7.980$	&	$86.073$ \\ 
HGF	&	$66$	&	$103$	&	$2$	&	$0$	&	$2$	&	$0$	&	$2.443$	&	$20.694$	&	$58.727$	&	- \\ 
T-diff	&	$68$	&	$175$	&	$6$	&	$0$	&	$6$	&	$0$	&	$1.245$	&	$13.475$	&	$18.790$	&	$14.103$ \\ 
HIV-1	&	$136$	&	$321$	&	$8$	&	$0$	&	-	&	-	&	$28.274$	&	-	&	$270.617$	&	- \\
\hline
\end{tabular}
\caption{An overview of the networks and a comparison of the two methods (TTC and SMC). Symbol `-' means that the method failed to finish the computation within five hours.}
\label{tab:overview}
\end{table*}

Our temporary target control method, described in Algorithms~\ref{alg:ttc} and~\ref{alg:verify}, 
is implemented in the tool ASSA-PBN~\citep{MPSY18} based on the model checker MCMAS~\citep{MCMAS} 
to encode Boolean networks into the efficient data structure BDD. 
All the experiments are performed on a high-performance computing (HPC) platform, which contains CPUs of Intel Xeon Gold 6132 @2.6 GHz.

As discussed in the introduction, both our method (TTC) and the stable motif-based control (SMC)~\citep{ZA15} 
focus on temporary target control of asynchronous Boolean networks. 
We apply our method on several real-life biological networks and compare its performance with SMC. 
Here we give a brief description on the networks. 
An overview of the networks can be found in Table~\ref{tab:overview}.

\begin{itemize}
\item The myeloid differentiation network is designed to model myeloid differentiation 
from common myeloid progenitors to four cell types, 
including megakaryocytes, erythrocytes, granulocytes and monocytes~\citep{KMST11}. 

\item The apoptosis network consists of necessary pro-apoptotic and anti-apoptotic pathways to 
capture decision-making on cell survival or apoptosis~\citep{TC09}. 

\item The cardiac gene regulatory network integrates major genes that play important roles 
in early cardiac development and  FHF/SHF determination~\citep{HGZKK12}. 

\item The ERBB receptor-regulated G1/S transition protein network combines ERBB signalling with G1/S transition of the mammalian cell cycle to identify new targets for breast cancer treatment~\citep{SFLK09}.

\item The HSPC-MSC network describes intercommunication pathways 
between hematopoietic stem and progenitor cells (HSPCs) 
and mesenchymal stromal cells (MSCs) in bone marrow (BM)~\citep{EMMP16}.

\item The PC12 cell network models temporal sequence of protein signalling, transcriptional responses and subsequent autocrine feedbacks~\citep{OKB16}.  

\item The network of hematopoietic cell specification covers major transcription factors and signalling pathways for lymphoid and myeloid development~\citep{COOAD17}.

\item The bladder cancer network allows us to identify deregulated pathways and their influence on bladder tumourigenesis~\citep{RRC15}.  

\item The MAPK network is constructed to study MAPK responses to different stimuli and their contributions to cell fates~\citep{GCT13}. 

\item The model of HGF-induced keratinocyte migration captures the onset and maintenance of hepatocyte growth factor-induced migration of primary human keratinocytes~\citep{SNB12}. 

\item The Th-cell differentiation network models regulatory elements and signalling pathways controlling Th-cell differentiation~\citep{NCCT10}.  

\item The HIV-1 network models dynamic interactions between human immunodeficiency virus type 1 (HIV-1) proteins 
and human signal-transduction pathways that are essential for activation of CD$4+$ T lymphocytes~\citep{ODRS14}. 
\end{itemize}

\noindent
\textbf{Attractors of the networks.}
Before the computation of target control, attractors are identified 
with our decomposition-based attractor detection method~\citep{MPQY17b} and SMC, respectively.  
Our method identifies all the exact attractors (the number of states and the structures for both singleton and cyclic attractors) introduced in Section~\ref{ssec:attractor}, 
while SMC identifies exact singleton attractors and quasi-attractors, which correspond to cyclic attractors. 
A quasi-attractor can be considered as a superset of an attractor: 
the values of oscillate nodes in the corresponding attractor are not specified in a quasi-attractor. 
Columns~$4$-$5$ and~$6$-$7$ of Table~\ref{tab:overview} show the number of attractors computed by the two methods. 
Most of the attractors identified by the two methods are the same except for 
the cyclic attractor of the apoptosis network (marked in bold in Table~\ref{tab:overview}).    
SMC identifies its quasi-attractor, which consists of $64$ states, 
while the corresponding cyclic attractor has $56$ states. 
Columns $8$ and $9$ of Table~\ref{tab:overview} 
show the execution time for attractor detection.   
We can see that our attractor detection method is more efficient than SMC. 

\begin{figure*}
\centering
\includegraphics[width=.95\textwidth]{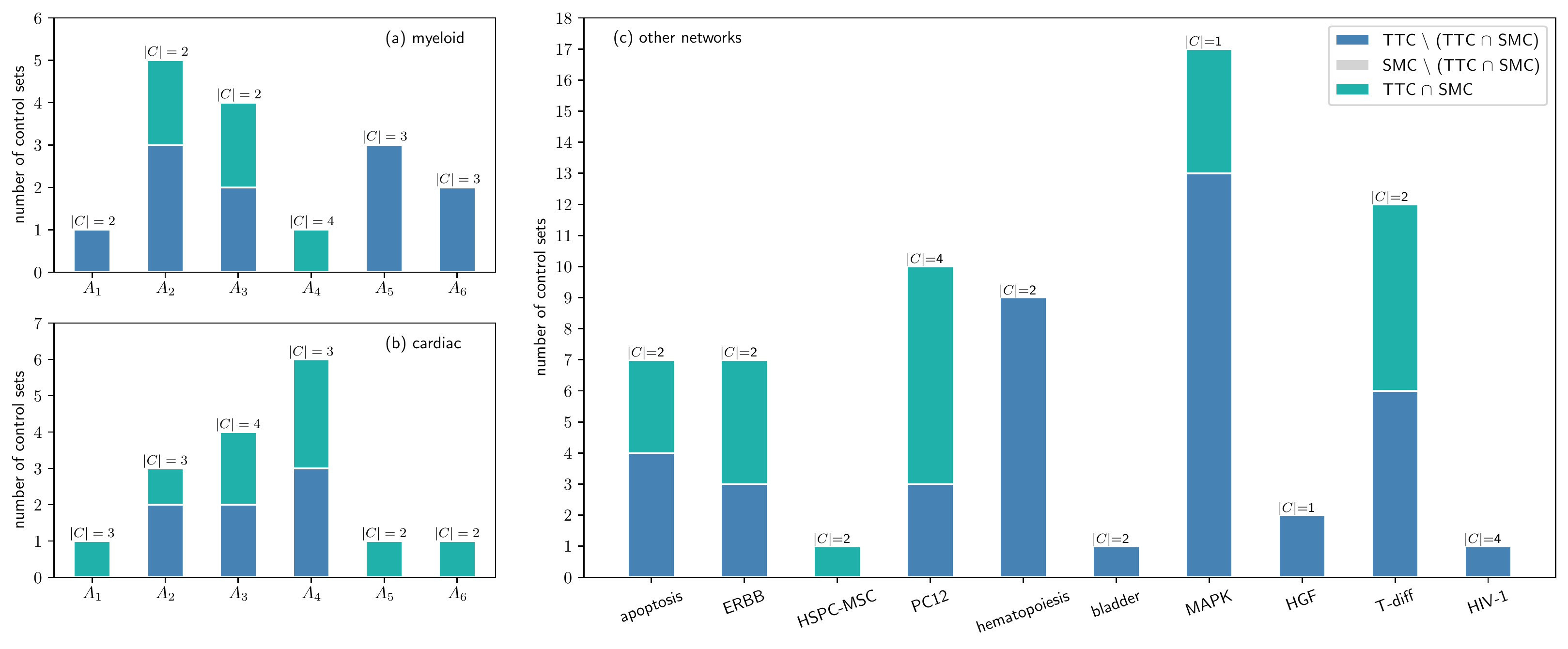}
\caption{An overview of the control results. 
The blue bars and grey bars represent the control sets that only appear in the results of TTC and SMC, respectively. 
The green bars denote the intersection of the two methods. }
\label{fig:networks}
\end{figure*}

\smallskip
\noindent
\textbf{Effectiveness.}  
We compute temporary target control for each attractor of the networks with TTC and SMC.  
Since neither of the methods guarantees the minimal control, 
they may find control sets of different sizes for one attractor. 
For comparison, we only consider the smallest control sets.

Fig.~\ref{fig:networks}~$(a)$ shows the number of smallest control sets for the myeloid differentiation network.    
The blue bars and grey bars represent the control sets that only appear in the results of TTC and SMC, respectively. 
The green bars denote the intersection of the two methods. 
The equation above each bar $|C|=k$ describes the size of control sets. 
For attractors $A_1$, $A_5$ and $A_6$, 
TTC identifies smaller control sets than SMC. 
Taking $A_5$, as an example, 
the minimal number of perturbations required by TTC and SMC is $3$ and $4$, respectively.
Since we only consider the smallest controls, 
SMC identifies no control sets of size $3$, thus we can only see the blue bar for this case.  
For other attractors ($A_2$, $A_3$ and $A_4$) of the myeloid network in Fig.~\ref{fig:networks}~$(a)$
and all the attractors of the cardiac network in Fig.~\ref{fig:networks}~$(b)$, 
two methods require the same number of perturbations, 
but our method has the potential to identify more solutions than SMC.

For the other networks listed in Table~\ref{tab:overview}, 
we summarise the number of control sets for one of the attractors in Fig.~\ref{fig:networks}~$(c)$. 
It shows that our method is able to identify smaller control sets than SMC 
for the bladder cancer network 
(SMC failed to compute results for hematopoiesis, HGF and HIV-1 networks).  
Our method also has the capability to provide more solutions, 
which may give more flexibility for clinical applications. 
Another interesting observation is that even for large networks, 
the number of perturbations is relatively small.

Now we use the myeloid differentiation network as an example to show 
the consistency of our results with biological conclusions in~\citep{KMST11}. 
This network consists of six attractors, four of which correspond to erythrocytes, megakaryocytes, monocytes and granulocytes. 
To realise the conversion to granulocytes ($A_5$ in Fig.~\ref{fig:networks} $(a)$) from any initial state, 
TTC needs to perturb C/EBP$\alpha$, PU.1, together with one of the nodes in $\{$cJun, EgrNab, Gfi1$\}$. 
It has been verified that coordinated overexpression of C/EBP$\alpha$ and PU.1 is required 
for the convergence to GM lineage (granulocytes and monocytes)~\citep{KMST11}. 
One more control node in $\{$cJun, EgrNab, Gfi1$\}$ helps to further distinguish granulocytes from monocytes. 

\smallskip
\noindent
\textbf{Efficiency.}  
The last two columns of Table~\ref{tab:overview} summarise the execution time 
for computing temporary target control for all the attractors of the networks. 
We can see that our method is more efficient than SMC for most of the cases. 
SMC failed to finish the computation for three networks (hematopoiesis, HGF, and HIV-1) within five hours. 
For the hematopoiesis network, SMC failed in the identification of stable motifs, 
which has been pointed out to be the most time-consuming part of SMC~\citep{ZA15}. 
The reason could be that the number of cycles and/or SCCs in its expanded network is computationally intractable. 
For the HGF-induced keratinocyte migration network, 
SMC is blocked in the optimisation of stable motifs due to that this network has $19$ stable motifs and
most of the stable motifs contain more than $16$ nodes. 
SMC failed to construct the expanded network representation for the HIV-1 network 
because some of its Boolean functions depend on many parent nodes ($K \geq 10$). 
Detailed discussion on the complexity of SMC can be found in~\citep{ZA15}.
The efficiency of our method is influenced by not only the network size, 
but also the number of attractors and the number of required perturbations. 
The results show that our method is quite efficient and scales well for large networks.

\section{Conclusion}\label{sec:conclusion}
In this work, we have developed a temporary target control method for asynchronous Boolean networks to identify a set of nodes, 
the temporary perturbation of which can drive the network from any initial state to the desired target attractor. 
We have evaluated our method on various biological networks to demonstrate its efficacy and efficiency. 

We compared our method with SMC, a promising method to solve the same control problem. 
SMC explores both structures and Boolean functions of Boolean networks, and is potentially more scalable for large networks.
In contrast, our method is essentially based on the dynamics of the networks,
and it will suffer the state space explosion problem for networks of several hundreds of nodes.
We believe that these two methods complement each other well.
In the near future, we aim to find a way to combine the strengths of both methods
by simultaneously exploring network structure and dynamics
to achieve more efficient computational methods
for the control of large biological networks.

\begin{acks}
This work was partially supported by the project SEC-PBN funded by University of Luxembourg
and the ANR-FNR project AlgoReCell ({\sf INTER/ANR/15/11191283}).
\end{acks}

\balance
\bibliographystyle{ACM-Reference-Format}
\bibliography{control}

\end{document}